\documentclass[journal]{IEEEtran}
\usepackage{cite}
\ifCLASSINFOpdf
\else
\fi

\usepackage{bbm}
\usepackage{subfigure}
\RequirePackage{graphicx}
\RequirePackage{amssymb,amsmath,bm}
\RequirePackage{textcomp}
\RequirePackage{booktabs}
\usepackage{enumitem}
\usepackage{url}

\begin{document}
\title{Audio-Visual Speaker Diarization: Current Databases, Approaches and Challenges}

\author{
        Victoria~Mingote,        
        Alfonso~Ortega,
        Antonio~Miguel,
        and~Eduardo~Lleida
\thanks{The authors are with ViVoLab, Arag\'{o}n Institute for Engineering Research (I3A), University of Zaragoza, Spain (e-mail: \{vmingote,ortega,amiguel,lleida\}@unizar.es.)}% <-this % stops a space
}

\maketitle

\begin{abstract}
Nowadays, the large amount of audio-visual content available has fostered the need to develop new robust automatic speaker diarization systems to analyse and characterise it. 
This kind of system helps to reduce the cost of doing this process manually and allows the use of the speaker information for different applications, as a huge quantity of information is present, for example, images of faces, or audio recordings.
Therefore, this paper aims to address a critical area in the field of speaker diarization systems, the integration of audio-visual content of different domains. 
This paper seeks to push beyond current state-of-the-art practices by developing a robust audio-visual speaker diarization framework adaptable to various data domains, including TV scenarios, meetings, and daily activities.
Unlike most of the existing audio-visual speaker diarization systems, this framework will also include the proposal of an approach to lead the precise assignment of specific identities in TV scenarios where celebrities appear.
In addition, in this work, we have conducted an extensive compilation of the current state-of-the-art approaches and the existing databases for developing audio-visual speaker diarization. 
%
%This paper aims to advance the field of speaker diarization by improving the integration of audio-visual content and proposing how to address the identity assignment in TV scenarios using the same framework. 
%
%This research will have far-reaching implications for applications ranging from information retrieval to enhancing user experiences in various audio-visual contexts. 

\end{abstract}


% \fi
\IEEEpeerreviewmaketitle

\section{Introduction}

\begin{figure*}[h]
    \centering
    \includegraphics[width=0.95\linewidth]{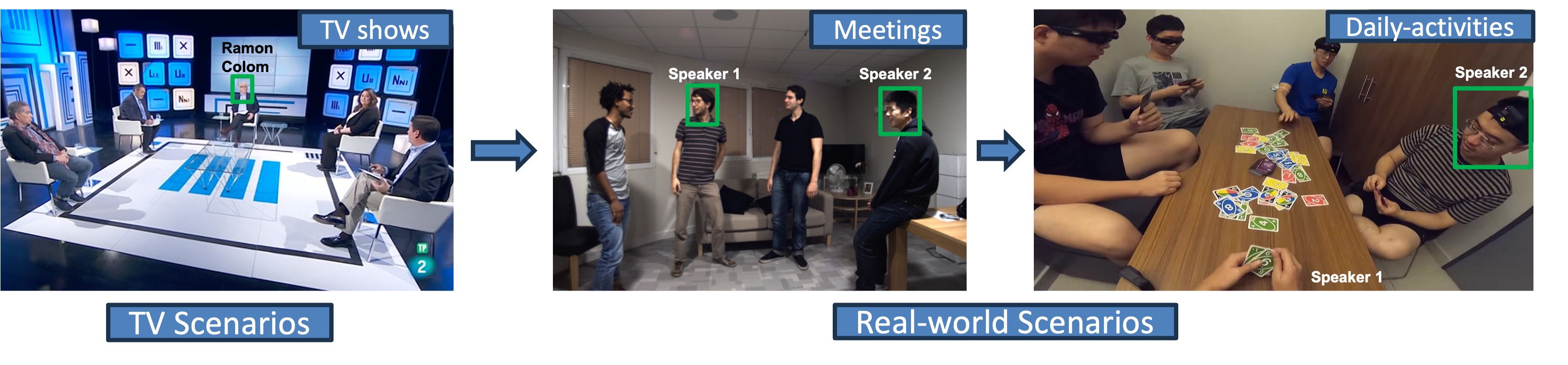}
    \caption{Challenge of realising a framework for enhancing the audio-visual speaker diarization and identification tasks across different scenarios.} 
    \label{fig1}
    %\vspace{-0.4cm}
\end{figure*}

%\newpage
\IEEEPARstart{W}{ith} the exponential growth of technological devices, most of us have a mobile phone, tablet or laptop which has profoundly changed the way to interact, communicate and learn. 
The rise of these devices with cameras and microphones has also promoted the generation of a large amount of new real-world audio-visual content to be processed such as meetings or daily activities from a first-person or egocentric point of view. 
Moreover, this technological expansion has motivated the digitalization of television (TV) repositories, which are composed of historical documents that need to be preserved. 
All this available audio-visual content could be beneficial for developing systems for different applications, as a huge quantity of information is present, for example, images of faces, scenes with different objects or audio recordings. 
Therefore, to increase the value of the audio-visual content, enhance the user experience and benefit society, this content has to be analysed and characterised, which is very expensive to do manually. 
For this reason, the development of new efficient tools is needed to process and extract relevant information automatically from this data, such as the segments where people speak, which is known as speaker diarization, the identities or the gender of the people present. 
This fact has led to a wide spread of systems based on artificial intelligence (AI) algorithms such as deep neural networks (DNNs) to address speaker diarization and identification tasks for different domains of audiovisual content \cite{mingote2021vivolab,mingote2022multimodal}. 
Hence, the main goal of this paper is to establish a robust automatic framework that allows the creation of systems that process the audio-visual content from different domains to obtain the speaker diarization information.
Moreover, the identity assignment or person recognition on top of this diarization information in TV scenarios is a task little studied in recent literature by applying novel AI algorithms and is of enormous interest for indexing and information retrieval in TV repositories.
Therefore, this paper also presents how this process could be included.

Speaker diarization consists of extracting the information of "who speaks when" in an audio or audio-visual recording.
The outcomes are the number of speakers and the assignment of each speech segment to a speaker cluster. 
It is a key task for many speech technologies such as automatic speech recognition (ASR), speaker verification (SV) and dialogue monitoring in different multi-speaker scenarios, including television (TV), meetings, and daily activities audio-visual recordings. 
%
%Traditionally, 
Early state-of-the-art systems developed to process the audio-visual content were focused on only the audio information to carry out the speaker diarization since audio-only-based approaches had been widely deployed to address the speaker diarization task. 
Using only audio content makes the task easier since the processing steps are fewer than in the case of employing visual information.
Audio speaker diarization usually employs a multi-stage framework composed of voice/speech activity detection (VAD/SAD), speaker embedding extraction \cite{xvector,dawalatabad21_interspeech}, and then clustering \cite{landini2022bayesian}.
However, whether there is uncertainty in the content, this kind of system still raises some issues with overlapping speech and ambiguous audios, such as those with background noise, music or low speaker similarity, while addressing this task.
Recent end-to-end (EEND) approaches merge all stages in audio speaker diarization systems \cite{Fujita2019Interspeech}.
These approaches attempt to solve these disadvantages, but several problems remain in complex environments such as when the number of speakers is large, or there is far-field and highly overlapped speech.

To enhance speaker diarization, researchers have found that the integration of visual information available in the audio-visual content is of paramount significance \cite{el2014audiovisual,shahabaz2024increasing}.
Since the human perception uses not only acoustic information but also visual information to reduce speech uncertainty.
Indeed, visual data, encompassing facial expressions, lip movements, and spatial positioning, adds a layer of context that complements audio data, providing information invaluable in scenarios characterized by the aforementioned problems. 
As audio-visual matching can improve time accuracy by merging speech activity with mouth/lip movement. 
Modality fusion can also act at the speaker similarity level, by merging audio and face embeddings. 
This synergy between audio and visual cues also empowers diarization systems to achieve a more complete understanding of content, leading to heightened precision in time and accuracy.

On the other hand, apart from the speaker segmentation itself, another relevant aspect of the evolution of speaker diarization systems is the precise assignment of specific identities to each speaker when they are famous people or celebrities such as in the case of TV archives \cite{everingham2006hello,gay2014comparison,mingote2021vivolab,mingote2022multimodal}.
The integration of visual information substantially aids in this identification process, particularly in cases where audio cues alone might be ambiguous. 
This combination of audio and visual cues fosters a comprehensive representation of speakers, enriching the system's capability to assign accurate identities. 

Nowadays, most of the existent audio-visual speaker diarization systems have been developed following a similar framework with at least 4 processing steps which are person localization and tracking \cite{kilicc2017audio,qian2021audio,zhao2023audiorev}, embedding fusion or active speaker detection \cite{robi2024active}, and speaker diarization process \cite{park2022review,reviewsd}. 
In addition, in the case of applying the identity assignment task, an extra final block is included on top of the speaker diarization outcome.
However, the specific approaches for each modality and processing step employed are not always the same since each system has been implemented focusing only on a single specific domain data. 
This data is usually based on controlled scenarios (meetings, TV shows or movies) due to the lack of uncontrolled data. 
Nevertheless, in the last few years, several challenges and datasets focused on uncontrolled data from real-world scenarios have been published (uncontrolled meetings or daily activities). 
Hence, in an era characterized by the proliferation of audio-visual data in sectors ranging from education and entertainment to security and communication, the imperative to cultivate adaptable audio-visual diarization systems that transcend data domain boundaries is emerging as a crucial asset. 
Furthermore, pursuing cross-domain diarization solutions fosters the evolution of generalized algorithms and methodologies. 
This adaptability not only maximizes utility but also underscores the significance of comprehensive, inclusive solutions in our interconnected digital landscape. 
Therefore, the systems developed will have the potential to provide a more robust AI solution for speaker diarization and identification across different domains as shown in Fig.\ref{fig1}.

Motivated by the previous issues regarding data and the lack of a common framework, this work presents an extensive analysis of this complex task, and the main contributions are summarized as follows:
\begin{itemize}
    \item An overview of existing approaches for the tasks of audio-visual person localization and tracking, active speaker detection, speaker diarization, and person recognition.
    \item A comprehensive explanation of all available databases for each of the above tasks to create the final audio-visual speaker diarization systems.
    \item The introduction of a baseline system capable of processing audio-visual content from different domains where the combination of audio-visual person localization, tracking and active speaker detection gives as a result the information necessary for audio-visual speaker diarization task. 
\end{itemize}

This paper is laid out as follows. 
Section \ref{sec:core} presents a review of core tasks for audio-visual speaker diarization plus the task of person recognition.
The review of databases relevant to each of these tasks is presented in Section \ref{sec:databases}. 
%
%In Section 4 other alternative approaches are described.
%
Section \ref{sec:system} shows the audio-visual speaker diarization system description, followed by the experimental data employed in Section \ref{sec:data}. 
Finally, Section \ref{sec:results} discusses the obtained results and Section \ref{sec:conclusion} concludes the paper.

%\newpage
\section{Core Tasks for Audio-Visual Speaker Diarization}
\label{sec:core}
To tackle the difficult task of speaker diarization for audio-visual content, more narrowly defined tasks have been described in the literature.
These tasks are the following: audio-visual person localization and tracking and active speaker detection. 
The combination of these two tasks taking into account the long-term relationship in the video files produces the outcome of audio-visual speaker diarization.  
The description and the most important approaches used for each of them are detailed below.
Furthermore, an extra subsection is included to detail the main idea and approaches employed in case the task of audio-visual person recognition is to be carried out.

\subsection{Audio-Visual Person Localization and Tracking}
Firstly, the goal of the person localization task consists of identifying and capturing the spatial position of all likely speakers in the audio-visual scene at a given moment.
When this task involves continuous monitoring over time, it is known as person tracking and plays a crucial role in different tasks such as scene understanding, speech enhancement, or speaker diarization.

Initial research efforts in this field focused on the analysis of static and controlled environments with simple scenarios \cite{lathoud2005av16, carletta2005ami,nawaz2011pft,deleforge2015,Gebru_2015_ICCV_Workshops,gebru2014audio}. 
In the literature, many approaches have been developed to address this task and most of them follow the same philosophy.
This philosophy consists of using visual and audio measurements combined with their respective likelihoods to obtain the localization information \cite{qian2019multi,schymura2020audiovisual,liu2023labelled}. 
Likelihood models can be categorized as discriminative or generative \cite{qian2021audio}.
After using these models for feature extraction, an audio-visual tracker is applied to link the two sources of information for the final tracking.
Most trackers adopt statistical methods such as approaches based on the Bayesian framework.
%or a new emerging line of techniques such as differentiable Bayesian or Transformer-based methods. 

In recent years, the emergence of new deep learning approaches has substantially changed the way many computer vision tasks are tackled.
However, at first, very few works attempted to use these approaches for audio-visual person tracking due to the lack of sufficiently large datasets for training.
Motivated by this issue, recently, a simulated dataset \cite{zhao2024attention} and a realistic dataset on a manually labelled robotic platform \cite{avridataset} have been proposed to enable the use of deep learning techniques.
Among the various attempts using these approaches, the combination of Bayesian trackers and these new deep learning approaches as a substitute for visual or audio measurements has shown an improvement over existing approaches to cope with more realistic and complicated scenarios with stronger reverberation and noise conditions \cite{li2022multi,zhou2024rav4d}. 
Another research stream has emerged to replace traditional Bayesian trackers with differentiable Bayesian or Transformer-based methods \cite{zhao2024attention}.
Furthermore, more recently, streams of research have begun to pay attention to this task from an egocentric point of view \cite{northcutt2020egocom, donley2021easycom}.
The creation of these new large egocentric databases is beneficial for the development of novel deep learning approaches to audio-visual person localization and tracking \cite{Jiang_2022_CVPR,zhao2023audio,murdock2024self}.

\subsection{Audio-Visual Active Speaker Detection}
Closely related to the previous task of speaker localization, audio-visual active speaker detection consists of determining whether any of the visible people are speaking at a given time.
Hence, unlike the previous task, this task does not take into account the long-term temporal relationship of the speakers but focuses on the current instant.
%
%The algorithms to address this process are based on using the information 

Traditionally, the audio-visual active speaker detection task has focused on lip motion detection and its correlation with speech.
Pioneering research in this field \cite{cutler2000look} detects correlated audio-visual signals by combining hand-craft feature sets extracted for audio and video signals using a time-delayed neural network (TDNN), but this work assumes that only one person is speaking at a time.
Follow-up works \cite{everingham2006hello} addressed the task when multiple speakers could appear and speak relying solely on visual information and considered a simpler set-up focusing strictly on lip and facial gestures.
Later, it was found that the use of audio to supervise the learning of a video-based active speaker system outperformed lip-motion for detecting active speakers in multi-speaker scenarios \cite{columbia1,li2005cross,everingham2009taking}.
In \cite{columbia1}, spatio-temporal features and speak/non-speak labels obtained from directional sound information were employed to train a video classifier.
All these initial approaches were developed using task-specific datasets of less than one hour and recorded in controlled environments with the possibility of using directional sound information to address the task.
Nevertheless, the exponential growth of audiovisual content with mobile devices and streaming platforms, where directional audio information is often missing, reflected the need for new approaches to deal with this data. 
Hence, in recent years, new databases have been collected from diverse audiovisual platforms \cite{chakravarty2016cross,chakravarty2016active,brown2021face,park2024kmsav,lin2024voxblink}.
In addition, in order to develop more robust audio-visual active speaker detection systems, novel approaches have been applied to this task. 
Among them, researchers have proposed audio-visual synchronization as a proxy task to jointly model audio-visual activity in a self-supervised manner \cite{chung2017out,afouras2020self,tao2021someone}.

But certainly to overcome the lack of diverse and in-the-wild data, the AVA-ActiveSpeaker database was proposed in \cite{roth2020ava}, which has been a key milestone in getting much more attention paid to this task.   
Moreover, given the relevance and success of this database, several more have been collected in recent years to try to complete the different situations that can occur in in-the-wild scenarios \cite{kim2021look,alcazar2021maas,roxo2023wasd}.
The availability of these new large in-the-wild databases has motivated the implementation of promising alternatives to follow the same philosophy of jointly modelling the activity on the faces and the concurrent audio stream and training them to detect a speech-to-face synchronization.
Most of the audio-visual active speaker detection approaches in the literature address the in-the-wild scenarios problem by relying on two-stage models.
First, audio and visual features are extracted using 2D or 3D Convolutional Neural Networks (CNNs) \cite{alcazar2020active,carneiro2021favoa,kopuklu2021design,datta2022asd,wuerkaixi2022rethinking,sharma2022unsupervised,sharma2022cross,sharma2023audio}.
Then, these systems learn the temporal synchronization based on these audio-visual features using Recurrent Neural Networks (RNNs), Graph Neural Networks (GNNs) \cite{alcazar2022end,min2022learning} or self-attention \cite{wang2024loconet,jung2024talknce,radman2024net}.

%\newpage
\subsection{Audio-Visual Speaker Diarization}
Audio-visual speaker diarization is a more complex task than the two mentioned above.
This task aims to identify the people talking in a video document and to quantify their speaking time.
In fact, the use of the information obtained with these other two simpler tasks is beneficial to achieve greater accuracy and performance in complex scenarios that exist today.

Early works were based on applying audio-only pipelines to solve the audio-visual task, as these approaches using audio streams were more advanced and simpler.
Nevertheless, more than two decades ago it was shown that facial attributes and lip movements are closely related to speech \cite{yehia1998quantitative}.
Researchers then began to consider that the use of visual information, when available, could provide complementary information to audio signals.
Thus, several early audio-visual speaker diarization methods exploit audio and visual cues driven by the synergy between utterances and lip movements.
These initial approaches employed techniques such as mutual information (MI) \cite{fisher2000learning,garau2010audio,noulas2012} or canonical correlation analysis (CCA) \cite{hotelling1992relations,kidron2007cross,sargin2007audiovisual}.
On the other hand, modelling the correspondence between talking faces and speech signals was also introduced in several works \cite{friedland2009,friedland2010dialocalization,vallet2012multimodal,minotto2014,minotto2015,gebru2015audio,bost2015audiovisual}.
In most of these approaches, an early feature fusion of audio and video features is performed and after that, a classifier is employed. 
Unlike the previous works, which enforce spatial coincidence, \cite{el2014audiovisual,kapsouras2017multimodal} proposes to cluster the two features independently and then correlate them based on the temporal alignments between speech and face segments.

The methods mentioned so far work well in formal conversations between several interlocutors, for example in meetings, when participants take turns to speak by generating clean voice signals with frontal images of faces.   
Moreover, in such scenarios the participants are seated or static, and there are usually dedicated closed-field microphones and cameras for each participant.
These datasets initially used for research have evolved from audio-visual person localization and tracking databases towards databases with annotations for diarization of speakers, as will be demonstrated in the next section.
However, in informal scenarios, e.g. social events or daily activities, the situation is much more complex as distant microphones provide the audio signals and may be corrupted by environmental noise, reverberations, and several people may speak simultaneously.
Furthermore, people often wander around, turn their heads away from the sensors, are hidden by other people, suddenly disappear from the camera's field of view, and reappear later.

To extend the research to these more complex situations, new, more robust approaches and appropriate databases had to be developed to study how to solve the task in these cases.
One of the first approaches to this was the development of a spatiotemporal model combining multiple-person visual tracking with multiple speech-source localization using a Bayesian fusion model \cite{gebru2017audio,dhaussy2023audio}.

Although these approaches were a breakthrough in the research field of this task, existing audio-visual diarization datasets were mainly focused on indoor environments such as meeting rooms or studios, which are quite different from scenarios with videos in the wild such as movies, documentaries or TV shows.
These scenarios present more complicated acoustic conditions, diverse domains and in some cases completely off-screen speakers.
Therefore, in the last five years, many new datasets and challenges have been presented to study and address these issues with current data \cite{chung2020spot,xu2021ava,chen2022misp,he2022,misp2022,liu2022msdwild,kwak2024voxmm} and also to digitize old data from different TV repositories \cite{lleida2018albayzin,lleida2020albayzin,ortega2020albayzin,ortega2022albayzin,ortega2024albayzin,larcher2020allies,larcher2021speaker}.

In addition, with the advent of the deep learning era, new approaches have been introduced to address this task.
Early research efforts with this philosophy followed the same previous ideas of exploiting the cross-modal relationship between audio and video modalities, and then merging the two sources of information.
Two types of strategies to fuse the information have been employed: early-fusion and late-fusion methods.
The former consists of joining audio and video feature vectors extracted from a deep neural network (DNN) and applying a final fusion method \cite{maurice2018odessa,chung2019said,xia2020online,ding2020self}.
While the latter performs this fusion at the score level, so audio and image diarization outcomes are obtained independently and once these results are achieved, a score fusion is carried out \cite{india2018upc,porta2021gtm,luna2021gth,mingote2021vivolab,mingote2022multimodal}.

On the other hand, the great development also in recent years of the active speaker detection task has promoted the introduction of audio-visual synchronization techniques based on attention mechanisms created for this task as a crucial step for a better audio-visual speaker diarization.
These approaches are often combined with clustering on audio-visual pairs to obtain the final results of audio-visual speaker diarization \cite{sharma2022using,wuerkaixidyvise,fanaras2022audio}.
In addition, the increase in the amount of data available for this task has also allowed the use of techniques based on end-to-end models (EEND) where the diarization process is unified in a single step \cite{qiu2022visual,he2022,casanet2023,yang2023uncertainty,cheng2024multi}.

Of late, an important new paradigm has emerged in recent years: augmented reality (AR) glasses.
This new paradigm has created a major challenge that still requires extensive research to be addressed.
For example, the egocentric perspective of a participant in the conversation, where off-screen speech activity must be detected and the quality of recorded videos is low, is a gap to be solved \cite{grauman2022ego4d,somasundaram2023project,min2022intel,min2023sthg}.

Beyond all the advances made during the last few years, there is a lack of a robust audio-visual framework that can effectively deal with different data domains where each scenario suffers from diverse problems such as different background noises, occlusion, off-screen speakers, or unreliable detection.

\begin{table*}[th!]
     \caption{Audio-Visual Speaker Diarization Databases. These databases are the ones available with annotations to perform the diarization task. $*$ denotes databases that initially were not created for diarization but in several works after that have established a protocol to be used for this purpose.}
    \label{tab:table1}
    \centering
    \resizebox{0.975\textwidth}{!} {
    \begin{tabular}{c | c | c | c | c | c}
    %\toprule
    \hline    
    \multicolumn{1}{c}{\textbf{Databases}}&
    \multicolumn{1}{c}{\textbf{Year}}& 
    \multicolumn{1}{c}{\textbf{Language}}& 
    \multicolumn{1}{c}{\textbf{Scenarios}}& 
    \multicolumn{1}{c}{\textbf{\# Videos}}& 
    \multicolumn{1}{c}{\textbf{Duration}}\\
    %\multicolumn{1}{c}{\textbf{Key Features}}\\
    \cline{1-6}
    $$AV16.3 \cite{lathoud2005av16}*$$&$2004$&$$ $$& $$Meetings$$& $40$& $$  $$\\
    $$AMI \cite{carletta2005ami}*$$&$2005$&$$English$$& $$Meetings$$& $684$& $$100h$$\\
    $$REPERE \cite{giraudel2012repere}$$&$2012$&$$French$$& $$TV Shows$$& $-$& $$6h$$  \\
    $$ETAPE \cite{gravier2012etape}$$&$2012$&$$French$$& $$TV Shows$$& $6$& $$29h$$  \\
    $$MVAD \cite{minotto2014,minotto2015}$$&$2014$&$$Portuguese$$& $$Meetings$$& $24$& $$21m$$ \\
    %$$MediaEval \cite{poignant2015multimodal,bredin2016multimodal}$$&$2015/16$&$$Multiple$$& $$TV shows$$& $ $& $$ $$ \\
    $$AVASM \cite{deleforge2015}*$$&$2015$&$$Multiple$$& $$Meetings$$& $4$& $$  $$\\
    $$AVTRACK-1 \cite{Gebru_2015_ICCV_Workshops}*$$&$2015$&$$Multiple$$& $$Meetings$$& $4$& $$-$$\\
    $$AVDIAR \cite{gebru2017audio}$$&$2018$&$$Multiple$$& $$Meetings$$& $23$& $$27m$$ \\
    $$RTVE 2018 \cite{lleida2018albayzin}$$&$2018$& $$Spanish$$&$$TV Shows$$& $4$& $$6h$$ \\
    $$CAV3D \cite{qian2019multi}*$$&$2019$& $$ $$&$$Meetings$$& $20$& $$ $$ \\
    $$RTVE 2020 \cite{lleida2020albayzin}$$&$2020$&  $$Spanish$$&$$TV Shows$$& $54$& $$33h 21m$$  \\
    %$$ALLIES \cite{larcher2020allies}$$&$2020$& $$French$$&$$TV Shows$$& $332$& $$119 h$$  \\
    $$Vox Converse \cite{chung2020spot}$$&$2020$&$$English$$& $$Youtube videos$$& $526$& $$73h 48m$$\\
    $$EgoCom \cite{northcutt2020egocom}*$$&$2020$& $$English$$&$$Egocentric/Daily Activities$$& $28$& $$ $$\\
    $$AVA-AVD \cite{xu2021ava}$$&$2021$& $$Multiple$$& $$Youtube videos$$& $ 351$& $$29h 15m$$\\
    $$EasyCom \cite{donley2021easycom}*$$&$2021$& $$English$$& $$Egocentric/Daily Activities$$& $12$& $$6h$$\\
    $$VCPD \cite{brown2021face}*$$&$2021$& $$English$$& $$TV Shows, Movies$$& $42$& $$23h 54m$$\\
    $$RTVE 2022 \cite{ortega2022albayzin}$$&$2022$&  $$Spanish$$&$$TV Shows$$& $38$& $$25h$$  \\
    $$Ego4D \cite{grauman2022ego4d}$$&$2022$&$$English$$& $$Egocentric/Daily Activities$$& $766$& $$63h 50m$$\\
    $$MISP \cite{misp2022}$$&$2022$&$$Chinese$$& $$Daily Activities$$& $311$& $$141h $$ \\
    $$MSDWILD \cite{liu2022msdwild}$$&$2022$& $$Multiple$$&$$Daily Activities$$& $3143$& $$84h$$\\
    $$KMSAV \cite{park2024kmsav}*$$&$2023$& $$Korean$$&$$Youtube videos$$& $512$& $$150h$$\\
    %$$RTVE 2024 \cite{ortega2024albayzin}$$&$2024$&  $$Spanish$$&$$TV Shows$$& $ $& $$ $$  \\
    $$VoxMM \cite{kwak2024voxmm}$$&$2024$& $$Multiple$$&$$Youtube videos (12 domains)$$& $289$& $$109h$$\\
    $$MMCSG \cite{somasundaram2023project}$$&$2024$& $$Multiple$$&$$Egocentric/Daily Activities$$& $530$& $$26h 18m$$\\
    %$$VoxBlink\cite{lin2024voxblink}*$$&$2024$& $$Multiple$$&$$Youtube videos$$& $241170$& $$1670 h$$\\

    %\bottomrule
    \hline
    \end{tabular}}
%\vspace{-0.2cm}
\end{table*}

\subsection{Audio-Visual Person Recognition}
Over the past decades, the task of identifying the real name of a person appearing or speaking in audio-visual content has been given different titles, such as audio-visual person recognition, identity assignment, or person discovery.
This task is of great relevance for information retrieval and indexing applications, mainly in TV repositories.
Therefore, some of the challenges aforementioned in the previous section for the audio-visual speaker diarization have also introduced the need for labels to perform this additional task.

The first of this type of challenge was the REPERE Challenge for person multimodal recognition \cite{giraudel2012repere,kahn2012presentation,gravier2012etape}.
As part of this challenge, several works were proposed to identify people from previously generated face/speaker clusters using different sources to obtain the names, such as written or pronounced names in audiovisual content \cite{bredin2013integer,gay2014comparison,Poignant2016}.
Right after the previous one, the MediaEval person discovery challenges were launched \cite{poignant2015multimodal,bredin2016multimodal}.
Following the same philosophy, several works continued to move in the same direction of using written or pronounced names to develop the systems for these new challenges \cite{8049254,nishi2016tokyo}.
Finally, the last series of this kind of challenge was the RTVE Multimodal Diarization and Identity Assignment Challenges \cite{lleida2018albayzin,lleida2020albayzin,ortega2020albayzin,ortega2022albayzin,ortega2024albayzin}.
In this case, the different versions of this series of challenges provide additional audio-visual content of people appearing and speaking, allowing the systems to build person models to identify them in the challenge videos 
\cite{mingote2021vivolab,mingote2022multimodal,india2018upc,maurice2018odessa,porta2021gtm,luna2021gth}.
Furthermore, in a parallel research stream, for almost two decades, several work attempts have been carried out to detect characters and automatically label them in movies or TV series \cite{everingham2006hello,nagrani2017benedict,bost2020serial,sharma2022audio,korbar2024look}.

\section{Audio-Visual Databases}
\label{sec:databases}
In this section, we introduce the audio-visual databases created in the last two decades to address the tasks of audio-visual person localization and tracking, active speaker detection, and speaker diarization.
Furthermore, Table \ref{tab:table1} presents a summary of the most relevant features of each of the databases employed for audio-visual speaker diarization.
This table also includes the information from the databases for audio-visual person localization and tracking and active speaker detection that were not collected specifically for the diarization task but were adapted in later works to be used for this task.

\subsection{Audio-Visual Person Localization and Tracking}
This section presents the datasets initially created for the audio-visual person localization and tracking or general audio-visual tasks and initially used for these specific speaker tasks.
These datasets are:

\begin{itemize}[leftmargin=*]
    
    \item \textbf{AV16.3} \cite{lathoud2005av16}: Audio-Visual data were recorded in a meeting room with 16 microphones with a sampling rate of 16 kHz and 3 cameras with a sampling rate of 25 frames per second (fps). 
    This collection consists of real-world data that includes a wide variety of situations, from one to three speakers seated most of the time, to walking near the table.
    Over 40 different audio-visual sequences make up the dataset, but only a small subset of these sequences have annotated ground truth labels to assess audio-visual person localization and tracking.

    \item \textbf{AMI} \cite{carletta2005ami}: Augmented Multi-party Interaction corpus contains 100 hours of homemade videos of meetings. 
    This dataset was recorded for different multi-modal tasks and was initially used for audio-visual speaker tracking.
    To capture this data, 6 cameras at 25 fps and 8 microphones with a sampling frequency of 48 kHz were employed.
    Moreover, each video session is composed of groups of four people.

    \item \textbf{SPEVI} \cite{nawaz2011pft}: Surveillance Performance EValuation Initiative collection includes data from the stereo audio and cycloptic vision (STAC) sensors, which is composed of two microphones on a long bar with a camera in the center. 
    The dataset is composed of eight sequences recorded in different scenarios with two microphones with a sampling rate of 44,1 kHz and a camera with a sampling rate of 25 fps.
    In five of the sequences, only one person appears, while in the other three, there are four speakers. 

    \item \textbf{AVASM} \cite{deleforge2015}: Audio-Visual recordings made in real-world conditions with a binaural acoustic dummy head with 2 microphones plugged into its ears and a camera placed under the head. 
    The microphones have a sampling frequency of 44,1 kHz, and the camera captures the images at 25 fps.
    The four recorded sequences of this dataset contain up to two participants.

    \item \textbf{AVTRACK-1} \cite{Gebru_2015_ICCV_Workshops}: The AVTRACK-1 dataset was recorded using a dummy head equipped with 4 microphones with a sampling frequency of 44.1 kHz and a camera with a sampling rate of 25 fps. 
    This database is composed of four audio-visual sequences consisting of conversations of up to three speakers.
    Moreover, the speakers present in the sequences are either static or moving always looking at the camera.
    
    %\item Speaker Tracking with Kinect2 (S3A) \cite{woodcock2016presenting}

    \item \textbf{CAV3D} \cite{qian2019multi}: This dataset was recorded using Co-located Audio-Visual platform for 3D tracking.
    CAV3D contains recordings with higher reverberation and more complicated scenarios than the previous audio-visual datasets.
    To make these recordings, a microphone circular array was used with a camera with a sampling rate of 15 fps and 8 microphones with a sampling rate of 96 kHz. 
    This dataset contains 20 sequences varying in length from 15 to 80 $s$ and with different numbers of speakers, with a maximum of three in one recording.

    \item \textbf{EgoCom} \cite{northcutt2020egocom}: Egocentric Communications is the first dataset of this type of egocentric natural conversations.
    To create this dataset, low-cost head-worn glasses were used to record stereo audio near the ears and video between the eyes.
    The audio was recorded at a sampling rate of 44.1 kHz, and the video was captured at 30 fps. 
    EgoCom comprises 28 conversations between 34 diverse speakers, with each conversation having three participants and at least two of them wearing the glasses to record.

    \item \textbf{EasyCom} \cite{donley2021easycom}: Easy Communications is a world-first dataset designed to help mitigate the cocktail effect of an egocentric, multi-sensory worldview driven by augmented reality (AR).
    This dataset has 12 egocentric video sessions of conversations in a simulated noisy environment, each with 4-6 participants. 
    To record these sessions, one participant per session was provided with AR glasses headset with 6 microphones recording audio at a sampling rate of 48 kHz, and a wide-angle camera capturing video at a frame rate of 20 fps.

    \item \textbf{AVRI} \cite{avridataset}: Audio-Visual Robotic Interface dataset was recorded to create a large-scale dataset to promote research in deep learning-based audio-visual speaker localization and tracking for human-robot interaction applications.
    The AVRI dataset was recorded using a Kinect sensor and a four-channel circular microphone array with a sampling frequency of 16 kHz.
    Both devices were mounted on a real robot that could move around in a real reverberant indoor room with various furniture.
    To record this data, 6 female and 5 male speakers participated and 43 different video sequences were captured.
    
    %\item \textbf{RAV4D} \cite{zhou2024rav4d}: Radar-Audio-Visual Dataset for Indoor Multi-Person Tracking 

\end{itemize}

\begin{figure*}[h]
\hspace{-0.8cm}
\begin{minipage}[b]{0.225\linewidth}
  \centering
  \centerline{\includegraphics[width=3.75cm]{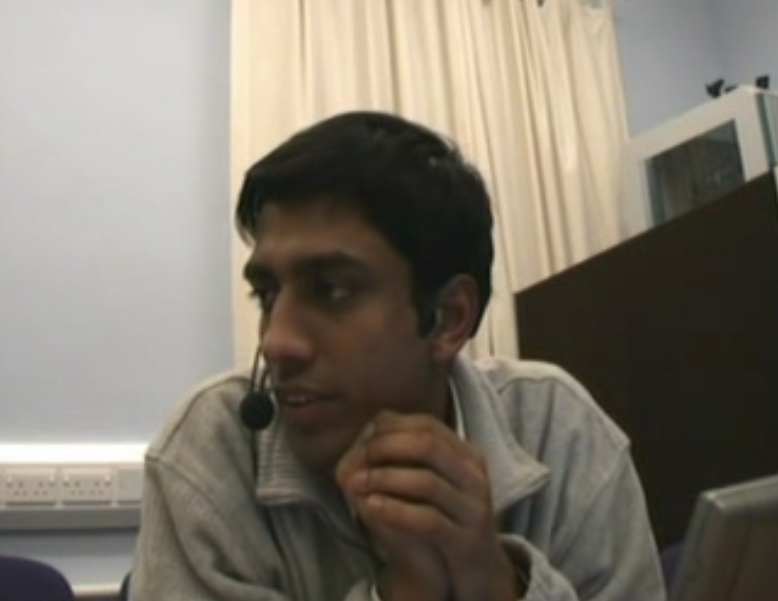}}
  \centerline{(a) AMI Meeting}\medskip
\end{minipage}
\hspace{-0.7cm}
\begin{minipage}[b]{0.225\linewidth}
  \centering
  \centerline{\includegraphics[width=4.2cm]{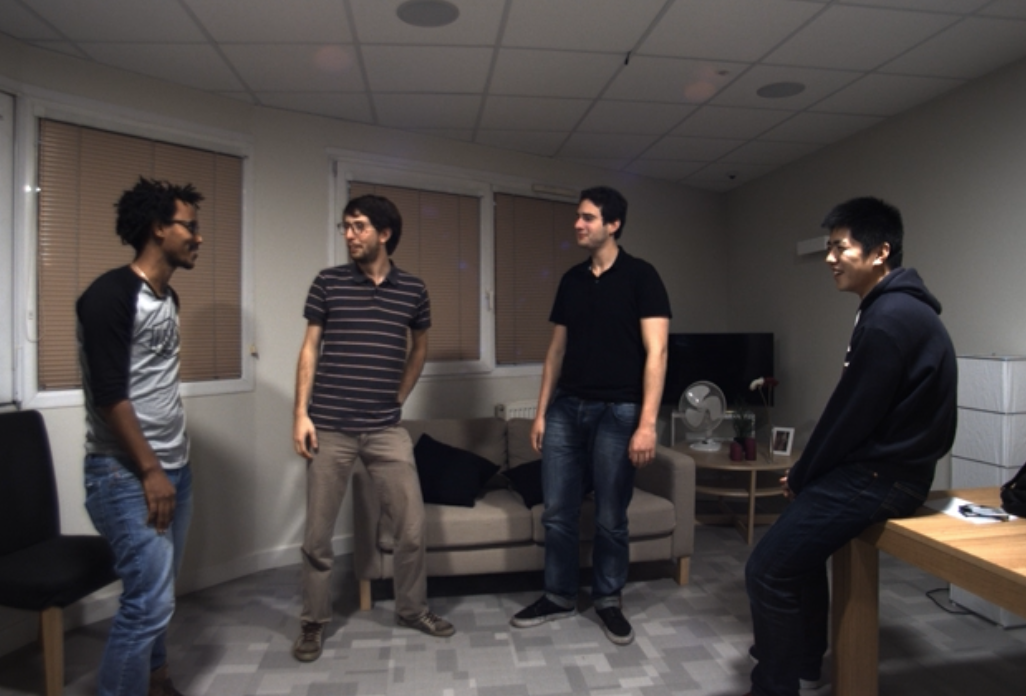}}
  \centerline{(b) AVDIAR}\medskip
\end{minipage}
\hspace{-0.2cm}
\begin{minipage}[b]{0.225\linewidth}
  \centering
  \centerline{\includegraphics[width=5.0cm]{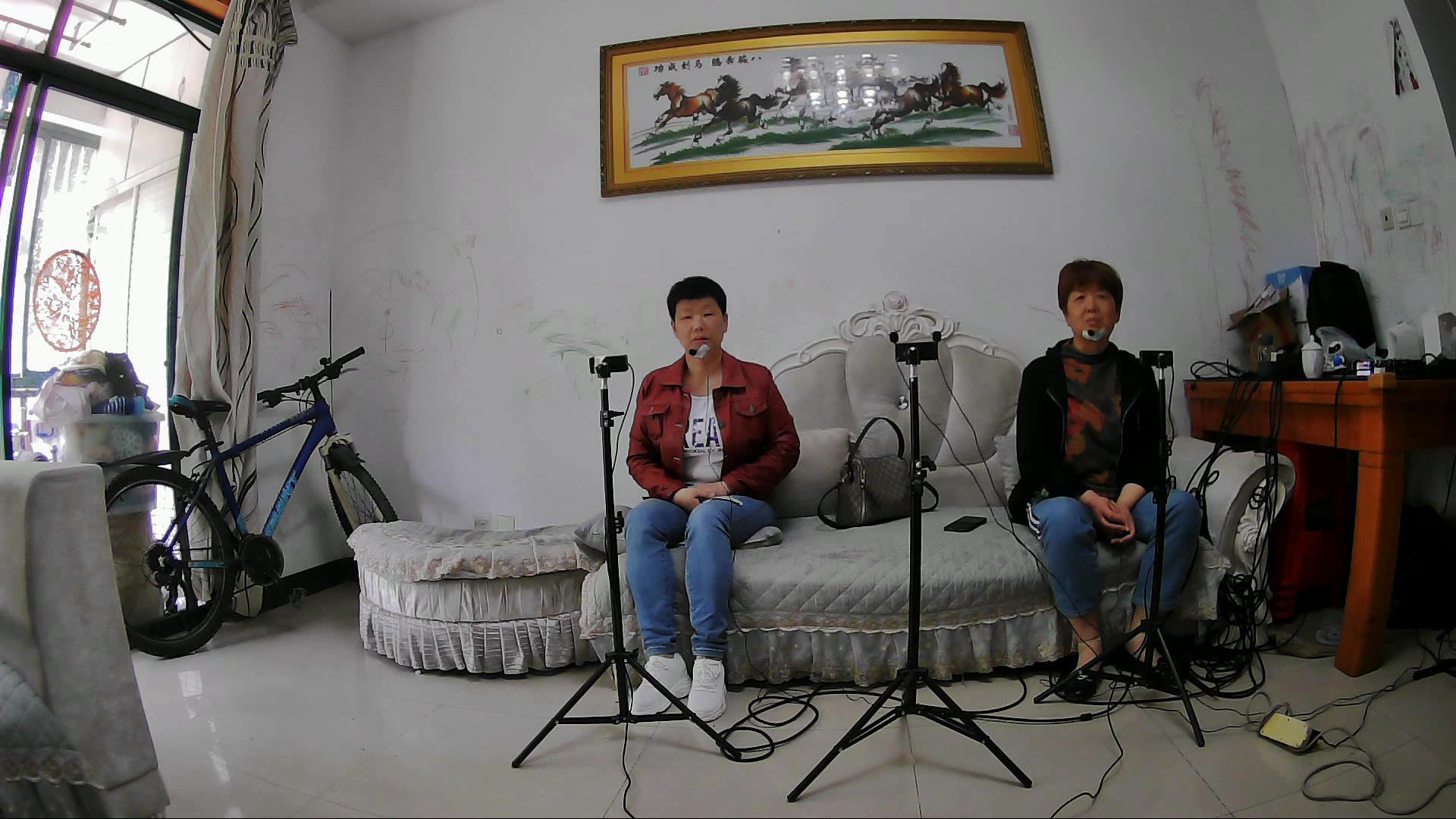}}
  \centerline{(c) MISP}\medskip
\end{minipage}
\hspace{0.1cm}
\begin{minipage}[b]{0.225\linewidth}
  \centering
  \centerline{\includegraphics[width=5.0cm]{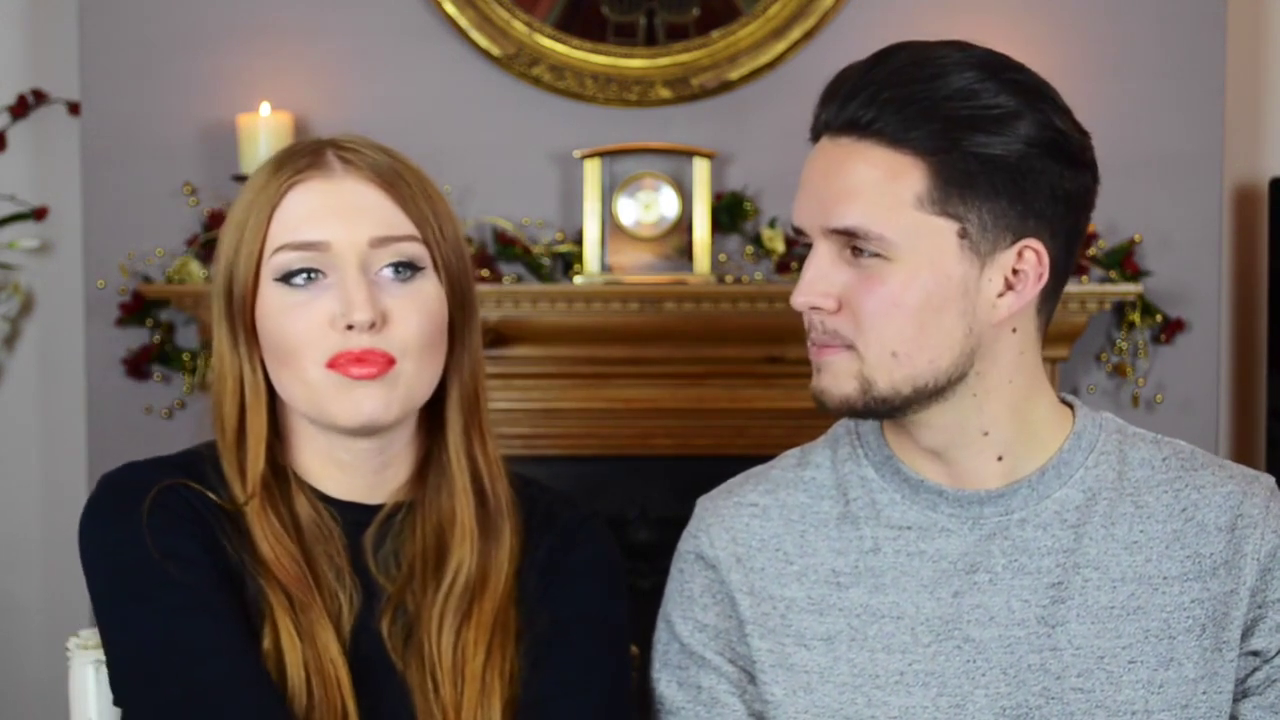}}
  \centerline{(d) MSDWILD}\medskip
\end{minipage}
\begin{minipage}[b]{0.225\linewidth}
  \centering
  \centerline{\includegraphics[width=4.7cm]{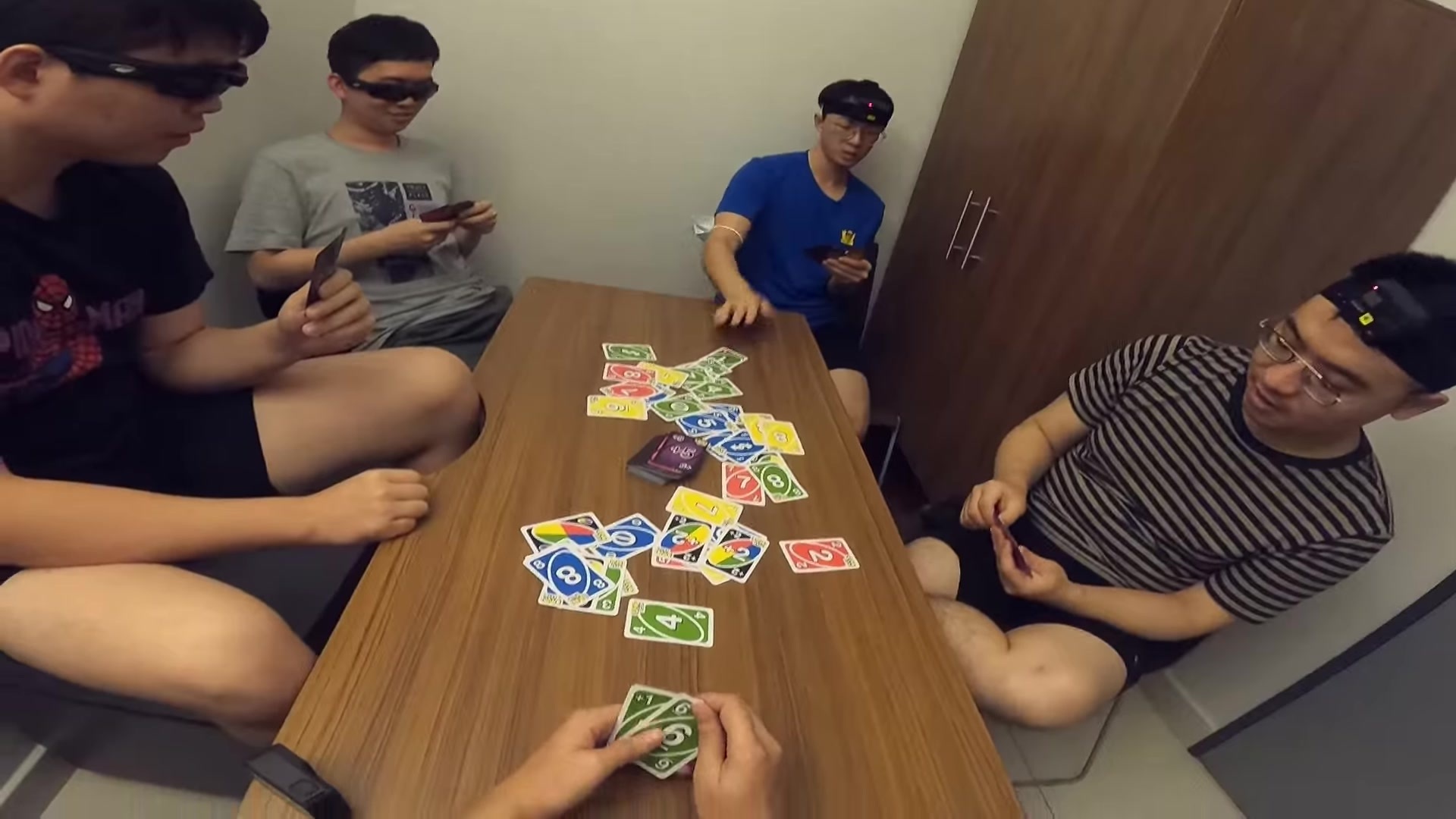}}
  \centerline{(e) Ego4D}\medskip
\end{minipage}
\hspace{0.1cm}
\begin{minipage}[b]{0.225\linewidth}
  \centering
  \centerline{\includegraphics[width=3.7cm]{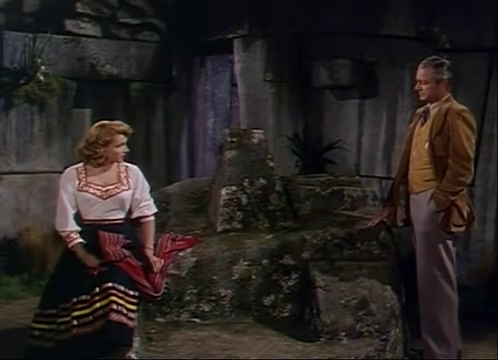}}
  \centerline{(f) AVA-AVD}\medskip
\end{minipage}
\hspace{0.05cm}
\begin{minipage}[b]{0.225\linewidth}
  \centering
  \centerline{\includegraphics[width=4.7cm]{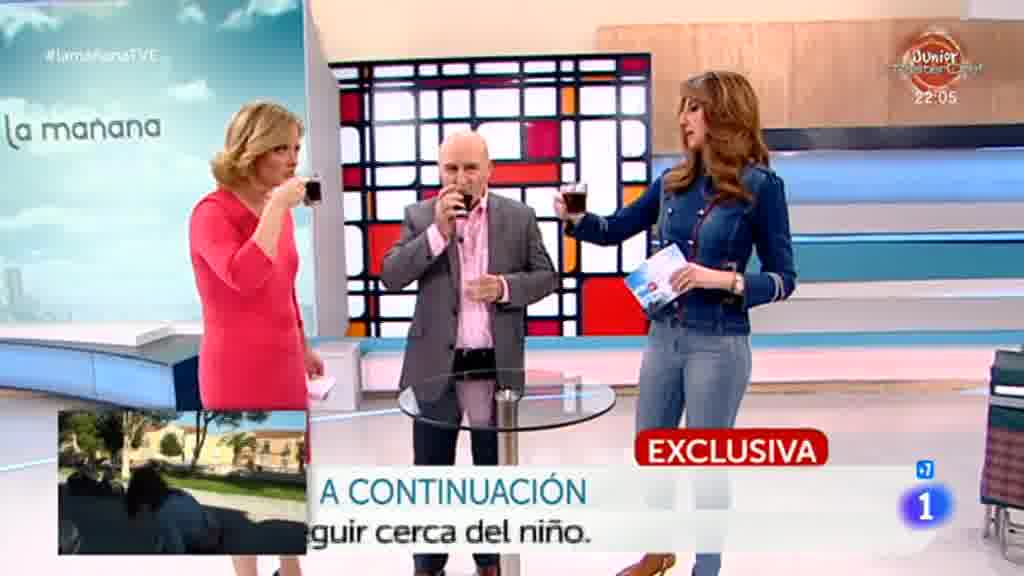}}
  \centerline{(g) RTVE2018}\medskip
\end{minipage}
\hspace{0.5cm}
\begin{minipage}[b]{0.225\linewidth}
  \centering
  \centerline{\includegraphics[width=4.7cm]{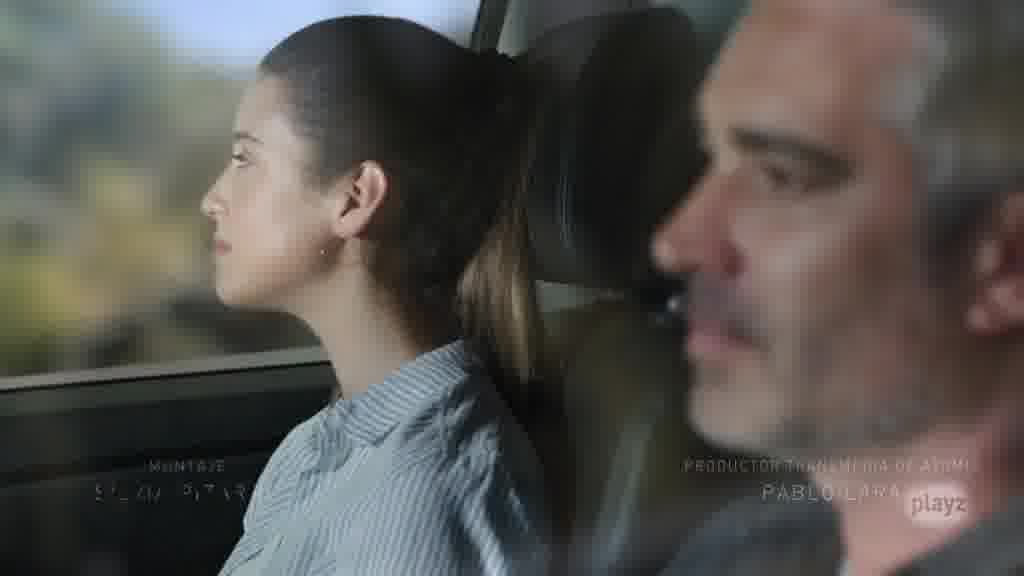}}
  \centerline{(h) RTVE2020}\medskip
\end{minipage}

\caption{Examples of Audio-Visual Speaker Diarization Databases. a-d) Speakers are always visible. e-h) Off-screen speakers are possible.} 
\label{fig2}
 %\vspace{-0.4cm}
\end{figure*}

\subsection{Audio-Visual Active Speaker Detection}
Unlike the audio-visual person localization and tracking databases, most publicly available datasets for active speaker detection have been collected from Internet videos.
Therefore, these datasets are more difficult and information on how these databases were recorded is more limited, as we will see below: 

\begin{itemize}[leftmargin=*]
    \item \textbf{Columbia} \cite{columbia1, chakravarty2016cross}: This dataset includes seven 30-minute thesis presentations and segments of a single 87-minute video of a panel discussion with 7 speakers at Columbia university.
    All these video sequences are available on YouTube.
    In this dataset, each speaker's upper body bounding boxes were annotated with speak and non-speak labels per frame. 
    
    \item \textbf{AVA-Active Speaker} \cite{roth2020ava}: Atomic Visual Action-Active Speaker detection dataset was the first large-scale database created for in-the-wild active speaker detection. 
    To create this dataset, 15-minute Hollywood movie clips from the 160 videos in the AVA v1.0 dataset \cite{gu2018ava} were annotated face bounding boxes with dense, spatio-temporal annotations of spoken activity. 
    These videos were recorded at different frame rates, 25-30 fps, and the number of speakers varies over time.
    In addition, this dataset contains movies from film industries worldwide, so the diversity of languages, face resolution and audible background noise, combined with the above, make this dataset a considerable challenge.
    
    \item \textbf{VPCD} \cite{brown2021face}: Video Person-Clustering dataset consists of full multi-modal annotations for main and supporting characters from videos of several widely watched TV shows (Friends, Sherlock, Buffy and TBBT) and 2 Hollywood movies (Hidden Figures and About Last Night).
    From all of these videos, a wide variety of characters were annotated with information in the bounding boxes about the faces that appeared in and their identities, along with information about which of them was speaking at any given moment.
    
    \item \textbf{ASW} \cite{kim2021look}: Active Speaker detection in the Wild dataset contains 212 videos randomly selected from the Voxconverse dataset \cite{chung2020spot}.
    In these videos, the face bounding boxes were detected in each frame and annotated with binary labels as active or non-active speakers.
    As a result, 30,9 hours were generated and annotated, including videos of news, debates, interviews, talk shows, etc.
    Hence, there are different types of background noise and visual problems in the videos.
    
    \item \textbf{Talkies} \cite{alcazar2021maas}: This dataset is composed of 4 hours and 10 minutes of manually labelled video data.
    The initial goal of collecting this dataset was to create a more diverse dataset including more multi-speaker situations, diverse characters, off-screen and noise.
    Therefore, the main source of videos employed was YouTube but, in this case, focusing on clips with this kind of data.
    This selection process results in 10.000 short clips of 1,5 seconds each.
    
    \item \textbf{WASD} \cite{roxo2023wasd}: Wilder Active Speaker Detection database contains 164 videos with a total of 30 hours of video annotations with face, body and speaker information.
    This dataset is divided into 5 categories with varying degrees of audio and facial quality.
    Moreover, each of these categories is balanced in terms of race, language and gender distribution.
    
    \item \textbf{KMSAV} \cite{park2024kmsav}: Korean Multi-speaker Spontaneous Audio-Visual speech corpus is the first corpus of spontaneous multi-speaker conversations in the Korean language.
    This dataset contains 512 YouTube videos with 150 hours of content transcribed and annotated with speaker labels.
    In addition, these videos have between 2 and 7 speakers in each and are grouped into 12 categories based on content and format.
    
    %\item \textbf{VoxBlink} \cite{lin2024voxblink}: A large-scale audio-visual speaker verification dataset  
\end{itemize}

\subsection{Audio-Visual Speaker Diarization}
This section presents the directly collected databases for the audio-visual speaker diarization.
Furthermore, some multimodal corpus created for person recognition or discovery, but including face and speaker diarization tasks, are also introduced below:

\begin{itemize}[leftmargin=*]
    \item \textbf{REPERE} \cite{giraudel2012repere,kahn2012presentation}: This corpus was created for a challenge to support the development of automatic systems for person multimodal recognition.
    In order to create this corpus, self-produced TV shows from 2 TV channels concerning debates and news were collected.
    In total, 6 hours of video were selected comprising TV shows with gradual difficulties in audio and video content. 
    
    \item \textbf{ETAPE} \cite{gravier2012etape}: This dataset contains 29 hours of TV broadcast from 3 TV channels with news, debates and entertainment.
    Moreover, the amount of overlapping speech is significantly higher than in other traditional news broadcast data, which encourages to focus on overlapping speech detection, diarization, and transcription tasks with this dataset.
    
    \item \textbf{MVAD} \cite{minotto2014,minotto2015}: Multimodal Voice Activity Detection dataset contains 31 single and multi-speaker audiovisual sequences recorded in an office environment.
    These sequences were captured using a Kinect sensor for the video signals at a sampling rate of 20 fps, and the audio was recorded with a uniform linear array of 8 microphones at a sampling frequency of 44,1 kHz.
    In this dataset, the recorded sequences range from 40 to 60 seconds long and contain one to three participants. 

    %\item \textbf{MediaEval} \cite{poignant2015multimodal,bredin2016multimodal}:
    
    \item \textbf{AVDIAR} \cite{gebru2017audio}: Audio-Visual speaker DIARization dataset was gathered and recorded to cover many multiple-speaker scenarios, such as static participants in front of the camera or with each other, moving participants,etc.
    A camera was attached to an acoustic dummy head to record this dataset.
    Video was captured at a sampling rate of 25 fps and microphone signals were recorded at 44,1 kHz.
    In this dataset, 27 video sequences containing one to four speakers were recorded.
    
    \item \textbf{RTVE 2018} \cite{lleida2018albayzin}: This dataset was created for a new challenge of the Albayzin evaluation series to address different issues such as the diversity of Spanish accents, overlapping speech, spontaneous speech, acoustic variability or background noise.
    In addition, on the output obtained from the speaker diarization, this challenge also introduces the task of identity assignment which is very important for certain applications such as Multimedia Indexing.
    For this indexing process, it is crucial to correctly assign as much of the content as possible, but it is also essential not to miss any speaker even if the amount of time it appears in the document is small.
    However, to build speaker models for the identity assignment task, enrollment data of the speakers is needed.
    Therefore, 39 characters with 10 images and a 20-second video of each character are available in this database.
    For the multimodal diarization task, 4 videos of 3 different TV shows of Radio Televisión Española (RTVE), the Spanish public radio and television, were annotated, so there are 6 hours of manually annotated audio-visual data.
    These videos contain a large number of speakers in each of them.
    
    \item \textbf{RTVE 2020} \cite{lleida2020albayzin,ortega2020albayzin}: Following on from the previous dataset, the RTVE 2020 Challenge was created as part of the 2020 edition of the Albayzin evaluations.
    This dataset is a collection of several broadcast TV shows covering different scenarios.
    To carry out this challenge, as an evaluation or test set, the database provides 54 videos of around 33 hours and 21 minutes of 9 new TV shows from RTVE different from those used in the RTVE2018 dataset.
    The development subset of the RTVE2020 database contains the RTVE 2018 database which is made up of four shows of around 6 hours.  
    Enrollment data is also provided for 161 characters with 10 images and a 20-second video of each character.

    \item \textbf{RTVE 2022} \cite{ortega2022albayzin}: This dataset continues with the same objective as the previous two challenges.
    The new evaluation data contains a set of 38 sequences from 9 different TV shows covering a variety of scenarios with a total duration of about 25 hours.
    More than 100 characters have been labeled with the identity of the real name and their corresponding enrollment files necessary for speaker identification are provided.
    The enrollment material consists of one or several audio files with more than 30 seconds of speech of each known character.

    %\item \textbf{RTVE 2024} \cite{ortega2024albayzin}:

    \item \textbf{VoxConverse} \cite{chung2020spot}: Audio-Visual Speaker Diarization in the Wild dataset includes 526
    videos from a range of multi-speaker acoustic environments, such as debates, panel discussions, celebrity interviews, etc.
    This dataset comprises approximately 73 hours of annotated video ranging from 22 seconds to 20 minutes in length.
    In addition, each video has on average between 4 and 6 speakers, and one video has 21 speakers.
    
    \item \textbf{AVA-AVD} \cite{xu2021ava}: Atomic Visual Action Audio-Visual Diarization in the Wild dataset was built upon the publicly available AVA-Active Speaker dataset \cite{roth2020ava}.
    This original dataset has 144 videos, from which 117 high-quality videos with diverse outdoor scenarios were selected to collect and annotate this new diarization database.
    Once this collection of videos was selected, they were divided into 5-minute clips, so the final number of sequences was 351.
    Furthermore, off-screen speakers were also annotated during the collection process, as this is a problem relevant to audiovisual speaker diarization that has not yet been extensively studied.
    
    \item \textbf{Ego4D} \cite{grauman2022ego4d}: Massive-scale Egocentric video dataset captures daily life activity around the world.
    This enormous dataset contains a huge amount of data and annotations for different audio-visual tasks, namely 3670 hours of video from 931 unique camera wearers.
    In terms of audio-visual diarization, there are 764 hours of videos relevant to this task.
    However, in the first available version only 572 videos with a duration of 48 hours were recorded.
    
    \item \textbf{MISP} \cite{chen2022misp,he2022,misp2022}: Multi-modal Information based Speech Processing dataset was developed as a challenge to tackle speech processing tasks, including audio-visual speaker diarization in everyday home environments.
    This dataset was recorded using two experimental setups: middle-field (1-1,5m) and far-field (3-5m).
    Videos in the former were captured using a linear microphone array of 2 microphones and a high-definition camera placed 1-1,5m away from the speakers, while videos in the latter were recorded using a linear microphone array of 6 microphones and a wide-angle camera located 3-5m from the speakers.
    The corpus is composed of 311 videos with a total duration of 141 hours.
    In each video, 2-6 speakers chat with each other or interact with the living room with TV background noise and varying lighting conditions.
    The total number of speakers in this dataset is 253 native Chinese participants, 98 males and 165 females.
    
    \item \textbf{MSDWILD} \cite{liu2022msdwild}: Multi-modal Speaker Diarization in the Wild dataset was collected from public YouTube videos, covering rich real-world scenarios and languages.  
    Compared to formal conversations, these real-world scenarios have three characteristics: frequently talking in turn, various head gestures, and several types of background noises.
    This dataset contains 3143 videos with 84 labelled hours. 
    Furthermore, this corpus was divided into 2 groups according to the number of participants in each video: few-talker set (two to four speakers) and many-talker set (more than four speakers).

    \item \textbf{VoxMM} \cite{kwak2024voxmm}: Vox Multi-modal Multi-domain corpus was collected from  YouTube videos to provide a more accurate representation of real-world scenarios.
    To address this challenge, 12 distinct conversation domains were included: daily conversations, commercials, entertainment, interviews, movies, lectures, politics, news, sports, documentaries, presentations, and remote meetings.
    In addition, a wide range of video lengths was covered, from 27 seconds to about 3 hours, which also allows simulating long-term conversations. 
    The total number of videos available in this dataset is 289, with a total number of 2.425 speakers. 

    \item \textbf{MMCSG} \cite{somasundaram2023project}: Multi-Modal Conversations in Smart Glasses dataset is composed of natural two-sided conversations.
    These conversations were recorded with Aria glasses, equipped with multiple microphones, cameras, and other sensors, and placed on people's head.
    Across the entire dataset, 138 speakers wearing the glasses were used to record 530 video clips.
    The audio sampling frequency was 48 kHz to record these clips, and the video sampling rate was 15 fps.

\end{itemize}

%\newpage
\section{Audio-visual System}
\label{sec:system}
After reviewing many of the existing databases and approaches, this section presents the audio-visual speaker diarization system established as a reference framework for processing content from different domains in this work.
Making the system robust and able to apply to different domains is not an easy task.
Hence, in the literature, no work addresses this complex task as each system has been implemented focusing only on a single specific data domain.
To develop the proposed framework, the baseline system available for Ego4D Challenge\footnote{\url{https://github.com/EGO4D/audio-visual}} has been employed as a starting point and some modifications have been applied.
As shown in Fig.\ref{fig3}, this system is composed of the following blocks and approaches for each block: 

\begin{figure}[h]
    \centering
    \includegraphics[width=1.0\linewidth]{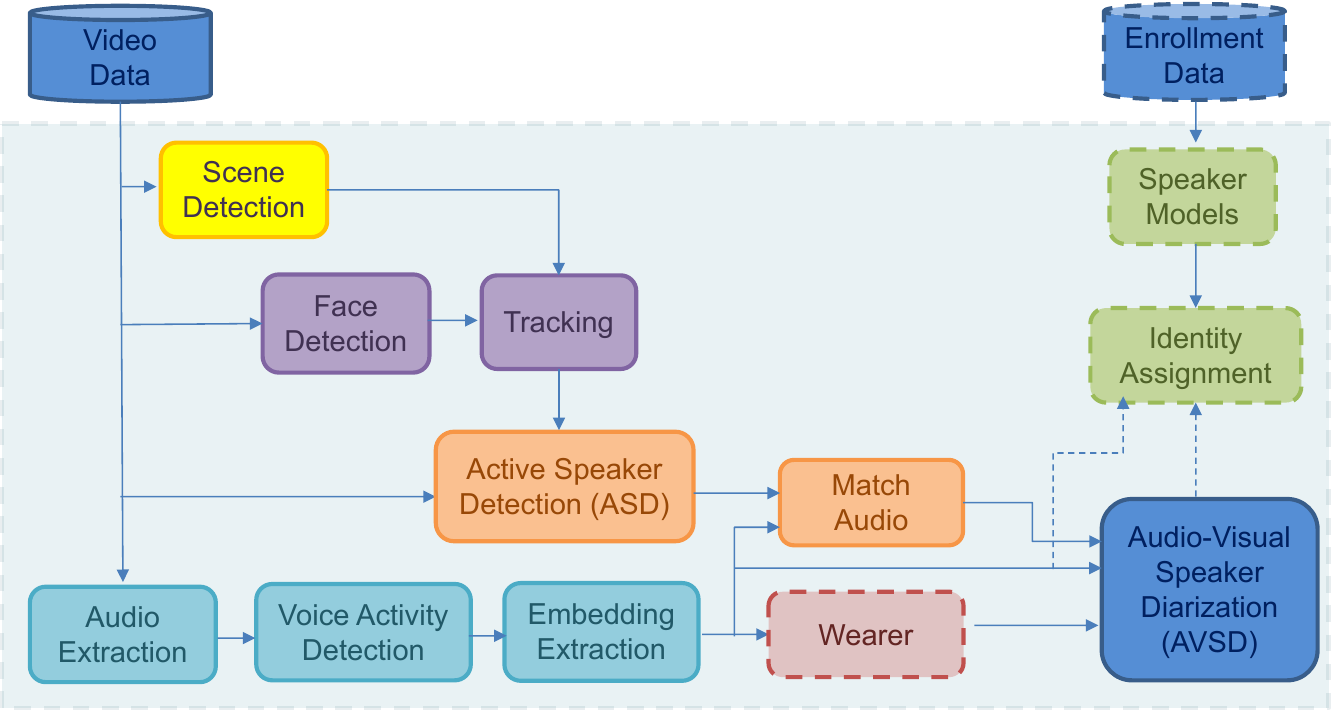}
    \caption{Audio-Visual Speaker Diarization system.} 
    \label{fig3}
    %\vspace{-0.4cm}
\end{figure}

\begin{itemize}[leftmargin=*]
    \item \textbf{Scene Detection}: In some audio-visual domains, videos are often composed of a huge variability in the content characteristics and constant changes of shots and scenes.
    Thus, to aid the tracking step, a scene change detection tool called PyScene\footnote{\url{https://www.pyscenedetect.readthedocs.io/en/latest/}} has been employed, as this tool efficiently detects these changes using the threshold-based detection mode.
    This detector finds areas where the difference between two subsequent frames exceeds a threshold value. 
    As a result, the detector provides the number of scenes and the start and end times of each scene.
    
    \item \textbf{Person Detection}: As a first step in the localization and tracking part, a person detector is applied on the videos to extract the information from the bounding boxes of each head and person appearing in a frame.
    The person detector is a Yolo-V3 detector \cite{redmon2018yolov3} that was trained with images from the Google OpenImage dataset \cite{kuznetsova2020open} and a fisheye image dataset \cite{kuznetsova2020open}.
    
    \item \textbf{Short-term Tracking}: After that, the features of all the detected heads are extracted using a ResNet-18 as the embedding network.
    In addition, a short-term tracker is employed to follow each detected head by expanding a set of trajectories.
    Trajectories may end when an occlusion happens or the person disappears from the field of view. 
    However, in this work, we have replaced the original OpenCV tracker implemented in C++ with a new one created in Python.
    Furthermore, this new tracker includes the information from the previous scene detection step.
    Entering this information allows us to restart the trajectory also when a change of scene occurs.
    
    \item \textbf{Global Tracking}: Once the previous short-term trajectories are generated along the whole video, a fast greedy algorithm is employed to group the different tracks from the same person.
    This process is made in two steps.
    First, the trajectories of each ID in a short-term time are linked together.
    After that, a comparison is made between embeddings of initially different IDs to join the same person that in the short-term tracking has been split with the assignation of several IDs.  
    
    \item \textbf{Active Speaker Detection}: The next step consists of classifying each person in each video frame as an active speaker or not. 
    To carry out this classification, the TalkNet approach\footnote{\url{https://github.com/TaoRuijie/TalkNet-ASD/}} has been applied.
    This approach is an end-to-end model where the input is the cropped faces and the corresponding audio, and using these resources it is decided whether the person is speaking in each frame. 
    Furthermore, this model has been trained using the AVA-Active Speaker dataset.
    As a result of this step, the above global tracking combined with the score prediction is stored to determine if each person is speaking or not.
    
    \item \textbf{Voice Activity Detection}: In addition to the previous steps, another stream is carried out to obtain speaker embeddings to complement the information on tracking and active speaker detection.
    First, however, a voice activity detector has been applied to segment the audio stream from the video content.
    To perform this audio segmentation, we have employed Silero-VAD\footnote{\url{https://github.com/snakers4/silero-vad}} which has shown great results in this task.
     
    \item \textbf{Embedding Extraction}: Once the audio segmentation is done, an embedding extractor is employed to obtain the speaker embeddings.
    In this work, these embeddings have been extracted using a modified ResNet18 trained on the VoxCeleb2 dataset using triplet loss.
    
    \item \textbf{Wearer Detection}: As mentioned above, some audio-visual domains have many off-screen speakers, so it is necessary to be able to detect speakers also out of the visual field of view. 
    Hence, in the framework of this work, an energy filter is used to detect the voice of invisible speakers (if they exist). 
    
    \item \textbf{Match Audio}: By combining information from global tracking, active speaker detection and wearer detection, matching is made between all this information to obtain the final speaker activities.
    This matching is performed by applying a threshold.
    
    \item \textbf{Audio-visual Speaker Diarization}: The last step of the framework consists of using all the information collected during the other steps to obtain the final speaker diarization outcomes. 
    Once these outcomes are obtained, the final performance is calculated with Diarization Error Rate (DER), which is a common metric to evaluate speaker diarization. 

     \item \textbf{Identity Assignment}: Finally, an additional step can be included if the assignation of identities is necessary.
     This work does not report these results but in the case of using it, two extra blocks are introduced in Fig.\ref{fig3}.
     First, the enrollment data available for identity identification is employed to generate speaker models.
     These models are compared with each audio embedding to associate them.
     Once this association is made, the unspecific IDs initially assigned in the previous step for diarization are replaced by the specific names of people.
     
\end{itemize}

%\newpage
\section{Experimental Data}
\label{sec:data}
%\subsection{Data}
In order to establish the proposed framework for audio-visual speaker diarization in Section \ref{sec:system}, different databases have been employed to evaluate it.
An example frame of each audio-visual speaker diarization database can be seen in Fig.\ref{fig2}. 

The initial idea was to use these 8 databases to establish the benchmark results with a common system for different domains.
However, when performing a preliminary analysis to study the organization and information available from each of them, we observed that in AMI Corpus \cite{carletta2005ami}, the data format is not compatible with the framework presented in this work.
In this database, each participant has his own video and two additional videos of the room from above and from the corner.
Furthermore, these videos do not have audio, so each session has a separate audio track.
Therefore, this data is not suitable for traversing the entire system.

On the other hand, the AVDIAR database \cite{gebru2017audio} also has several videos for each session, but in this case, all the videos are from different locations but with all scenes visible and audio available.
Hence, all 27 videos in this database have been evaluated with the proposed framework.
In addition, the other 6 databases have been employed to perform the experiments.
Of the MISP dataset \cite{chen2022misp}, only the development set composed of 10 videos has been used, as the reference files for the evaluation set are not publicly available.
In the case of the MSDWILD corpus \cite{liu2022msdwild}, 177 videos are available to evaluate the system in the worst case with many talkers in each video.
%using data from TV home scenarios.

The above databases are based on audio-visual content from TV home scenarios and meetings without off-screen speakers and continuous and abrupt shot changes.
Therefore, to cover a wider range of domains, 50 videos from the Ego4D validation set \cite{grauman2022ego4d} have been evaluated.
Moreover, the evaluation set of AVA-AVD database \cite{xu2021ava} with 18 videos of Hollywood movies has been used.
Finally, two challenges have been included to promote the need to develop better tools for the digitalization of TV repositories.
The RTVE 2018 Challenge \cite{lleida2018albayzin} only has 4 videos annotated for the multimodal diarization challenge, so those are the ones that have been evaluated.
From the RTVE 2020 Challenge \cite{ortega2020albayzin}, the evaluation set composed of 54 videos has been employed in this work.

%\vspace{-0.2cm}g
%\subsection{System configuration}

%\vspace{-0.2cm}
%\newpage
\section{Results}
\label{sec:results}
In this work, to evaluate the proposed framework, we have selected 7 databases which cover a wide range of different data domains.
To measure the performance of these experiments, the metric used is diarization error rate (DER) which is usually the reference metric employed in the diarization task.
This metric can be decomposed into the three terms of error: probability of misses (MISS), probability of false alarm (FA), and Speaker error (SPKERR).
The decomposition in these terms allows us to better analyze the results obtained and why they occur. 

Table \ref{tab:table2} presents 

\begin{table}[th!]
    \caption{DER Results on AVDIAR, MISP and MSDWILD databases \cite{gebru2017audio,chen2022misp,he2022,liu2022msdwild}.}
    \label{tab:table2}
    \centering
    \resizebox{0.475\textwidth}{!} {
    \begin{tabular}{c | c| c | c | c }
    %\toprule
    \hline    
    \multicolumn{1}{c}{\textbf{Database}}&
    %\multicolumn{1}{c}{\textbf{Subset}}& 
    \multicolumn{1}{c}{\textbf{DER\%}}&
    \multicolumn{1}{c}{\textbf{MISS\%}}&
    \multicolumn{1}{c}{\textbf{FA\%}}&
    \multicolumn{1}{c}{\textbf{SPKERR\%}}\\
    \cline{1-5}
    %$$AMI Meeting$$&$ $&$ $&$ $\\
    %\hline
    $$AVDIAR$$&$TBC$&$TBC$&$TBC$&$TBC$\\
    \hline
    $$MISP$$&$TBC$&$TBC$&$TBC$&$TBC$\\
    \hline
    $$MSDWILD$$&$TBC$&$TBC$&$TBC$&$TBC$\\
    %\bottomrule
    \hline
    \end{tabular}}
%\vspace{-0.2cm}
\end{table}

While in Table \ref{tab:table3},

\begin{table}[th!]
    \caption{DER Results on Ego4D, AVA-AVD, RTVE 2018 and 2020 databases \cite{grauman2022ego4d,xu2021ava,lleida2018albayzin,lleida2020albayzin,ortega2020albayzin}.}
    \label{tab:table3}
    \centering
    \resizebox{0.475\textwidth}{!} {
    \begin{tabular}{c | c | c | c | c  }
    %\toprule
    \hline    
    \multicolumn{1}{c}{\textbf{Database}}&
    %\multicolumn{1}{c}{\textbf{Subset}}& 
    \multicolumn{1}{c}{\textbf{DER\%}}&
    \multicolumn{1}{c}{\textbf{MISS\%}}&
    \multicolumn{1}{c}{\textbf{FA\%}}&
    \multicolumn{1}{c}{\textbf{SPKERR\%}}\\
    \cline{1-5}
    $$Ego4D $$&$TBC$&$TBC$&$TBC$&$TBC$\\
    \hline
    $$AVA-AVD$$&$TBC$&$TBC$&$TBC$&$TBC$\\
    \hline
    $$RTVE 2018 $$&$TBC$&$TBC$&$TBC$&$TBC$\\
    \hline
    $$RTVE 2020 $$&$TBC$&$TBC$&$TBC$&$TBC$\\
    %\bottomrule
    \hline
    \end{tabular}}
%\vspace{-0.2cm}
\end{table}

%\begin{table}[th!]
%    \caption{DER Results on AMI Meeting, AVDIAR, AVA-AVD, Ego4D, MISP and MSDWILD databases \cite{xu2021ava,grauman2022ego4d,chen2022misp,he2022,liu2022msdwild}.}
%    \label{tab:table2}
%    \centering
%    \resizebox{0.475\textwidth}{!} {
%    \begin{tabular}{c | c | c }
    %\toprule
%    \hline    
%    \multicolumn{1}{c}{\textbf{Database}}&
%    \multicolumn{1}{c}{\textbf{Subset}}&  
%    \multicolumn{1}{c}{\textbf{DER\%}}\\
%    \cline{1-3}
%    $$AMI Meeting$$&$$dev$$&$ $\\
%    $ $&$$test$$&$ $ \\
%    \hline
%    $$AVDIAR$$&$$dev$$&$ $\\
%    $ $&$$test$$&$ $ \\
%    \hline
%    $$MISP$$&$$dev$$&$ $\\
%    $ $&$$test$$&$ $\\
%    \hline
%    $$MSDWILD$$&$$dev$$&$ $\\
%    $ $&$$test$$&$ $\\
    %\bottomrule
%    \hline
%    $$Ego4D $$&$$dev$$&$ $\\
%    $ $&$$test$$&$ $ \\
%    \hline
%    $$AVA-AVD$$&$$dev$$&$ $\\
%    $ $&$$test$$&$ $ \\
%    \hline
%    \end{tabular}}
%\vspace{-0.2cm}
%\end{table}

%\begin{table}[th!]
%    \caption{DER Results on RTVE 2018 and 2020 databases \cite{lleida2018albayzin,lleida2020albayzin,ortega2020albayzin} with identity assignment.}
%    \label{tab:table1}
%    \centering
%    \resizebox{0.475\textwidth}{!} {
%    \begin{tabular}{c | c | c }
%    %\toprule
%    \hline    
%    \multicolumn{1}{c}{\textbf{Database}}&
%    \multicolumn{1}{c}{\textbf{Subset}}&  
%    \multicolumn{1}{c}{\textbf{DER\%}}\\
%    \cline{1-3}
%    $$RTVE 2018 $$&$$dev$$&$ $\\
%    $ $&$$test$$&$ $ \\
%    \hline
%    $$RTVE 2020 $$&$$dev$$&$ $\\
%    $ $&$$test$$&$ $\\
%    %\bottomrule
%    \hline
%    \end{tabular}}
%%\vspace{-0.2cm}
%\end{table}

%\vspace{-0.3cm}
%\newpage
\section{Conclusion}
\label{sec:conclusion}
In this paper, we have made  

room for improvement

%\newpage
\section*{Acknowledgment}
This work has been supported by the European Union’s Horizon 2020 research and innovation programme under Marie Skłodowska-Curie Grant 101007666; in part by MCIN/AEI/10.13039/501100011033 and by the European Union “NextGenerationEU” / PRTR under Grant PDC2021-120846-C41, and by the Government of Aragón (Grant Group T36\_22R).

\ifCLASSOPTIONcaptionsoff
  \newpage
\fi



%\bibliographystyle{IEEEtran}
%\newpage
\bibliographystyle{IEEEtran}

\bibliography{mybib}



%\newpage



%\vfill


\end{document}